\documentclass{ws-p8-50x6-00}
\input psfig.sty
\begin{document}

\title{ Deuteron Electromagnetic Properties with a Poincar\'e-Covariant Current
Operator within Front-Form Hamiltonian Dynamics }

\author{E. Pace}


\address{Dipartimento di Fisica, Universit\`{a} di Roma "Tor Vergata" and
Istituto Nazionale di Fisica Nucleare, Sezione Tor Vergata, Via della Ricerca
Scientifica 1, I-00133 Roma, Italy}

\author{G. Salm\`{e}}

\address{Istituto Nazionale di Fisica Nucleare, Sezione Roma I, P.le A. Moro
2, I-00185 Roma, Italy}




\maketitle

\abstracts{
Deuteron elastic and deep inelastic electromagnetic properties have been studied
within the front-form Hamiltonian dynamics, using a Poincar\'e-covariant current
operator. The deuteron elastic form factors are strongly sensitive to different
realistic $N-N$ interactions, while the relevance of different nucleon form
factor models is huge for $A(Q^2)$, weak for $B(Q^2)$ and negligible for the
tensor polarization. The possibility to gain information on the neutron charge
form factor from an analysis of $A(Q^2)$ has been investigated. The extraction of the neutron structure functions from the
deuteron deep inelastic structure functions at high $x$ is largely affected by
the use of our Poincar\'e-covariant relativistic approach instead of the usual
impulse approximation within an instant-form approach.}

\section{Introduction}
\indent  In Ref. \cite{LPS} we have constructed an electromagnetic current
operator for bound systems of interacting particles that satisfies Poincar\'e,
parity and time-reversal covariance, together with Hermiticity and current
conservation. This operator has been used to evaluate the deep inelastic
structure functions (DISF) in an exactly solvable two-body model in Ref.
\cite{LPS1}, and the deuteron elastic form factors (ff) in Refs.
\cite{LPS2,LPS3}. Our current operator is obtained within the front-form
Hamiltonian dynamics, starting from the free current in the Breit frame where
initial and final three-momenta of the interacting system are directed along the
spin quantization axis, $z$. In order to implement Hermiticity a term which
implicitely introduces two-body terms in the current has to be considered. In
the elastic case the continuity equation is automatically satisfied. By using
this covariant current, one gets rid of the well-known ambiguities, which in the
front-form approach plague the extraction of elastic and transition form factors
for spin $\geq 1$ systems, when the free current is considered in the $q^+=0$
reference frame. Indeed, in our approach the results for elastic and transition
form factors, as well as for deep inelastic structure functions, are independent
of the  matrix elements of the current which are employed in the calculation,
as it must be. Then the deuteron elastic ff and DISF can be safely calculated
from front-form deuteron wave functions. 

In this paper we report the results of our thorough investigation on the
sensitivity of deuteron ff to different realistic $N-N$ interactions and to
different nucleon form factor models \cite{LPS2}$^-$\cite{Evora}. Furthermore we
report our studies on the extraction of the neutron structure function,
$F_2^n(x)$, from the deuteron deep inelastic structure function, $F_2^D(x)$ and
on the relevance of our Poincar\'e-covariant relativistic
approach \cite{PSS,PSS1}. 

\begin{figure}[t]

\psfig{figure=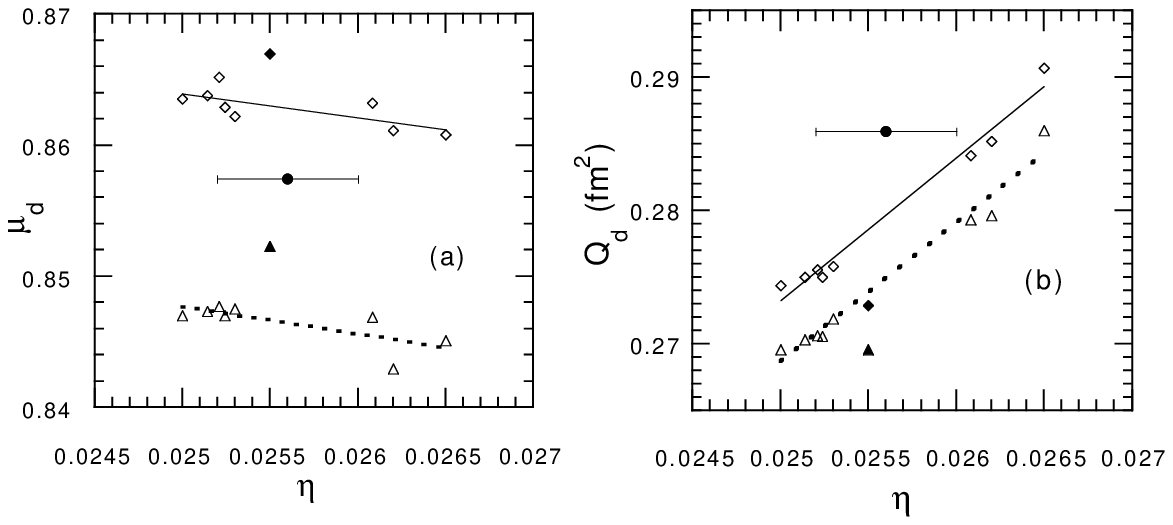,bbllx=11mm,bblly=215mm,bburx=0mm,bbury=280mm}
   
Figure 1.  { (a) Deuteron magnetic moment, $\mu_d$, vs the asymptotic
normalization ratio $\eta=A_D/A_S$, for different $N-N$ interactions. The full
dot represents the experimental values for $\mu_d$ and $\eta$; empty triangles
and diamonds correspond to the nonrelativistic and relativistic results,
respectively, while the solid and dashed lines are linear fits for these
results. Full triangle and diamond are the results of the $CD-Bonn$ interaction.
(b) The same as in (a), but for the deuteron quadrupole moment, $Q_d$ (After
Ref. \cite{LPS2}) } 

\end{figure}

\section{Deuteron Elastic Form Factors}
As a first application of our current operator for elastic electron scattering,
we determined \cite{LPS2} the deuteron magnetic and quadrupole moments, $\mu _d$
and $Q_d$, by adopting front-form deuteron wave functions corresponding to a
large set of realistic $N-N$ interactions ($RSC$, $Paris$ \cite{Par}, $Argonne$
$V14$ \cite{AV} and $V18$ \cite{AV18}, $Nijmegen-1$, $-2$, $-93$ and $Reid93$
\cite{Nijm}, $CD-Bonn$ \cite{CDBonn}). For both quantities a linear behaviour
against the deuteron asymptotic normalization ratio $\eta=A_D/A_S$ was found 
(see Fig. 1). Therefore, the usual disagreement between theoretical and
experimental results is largely removed and explicit contributions of dynamical
two-body currents  and isobar components in the deuteron should play a minor
role at $Q^2 \rightarrow 0$, using our Poincar\'e-covariant relativistic
framework \cite{LPS2}.

\begin{figure}[t]

\psfig{figure=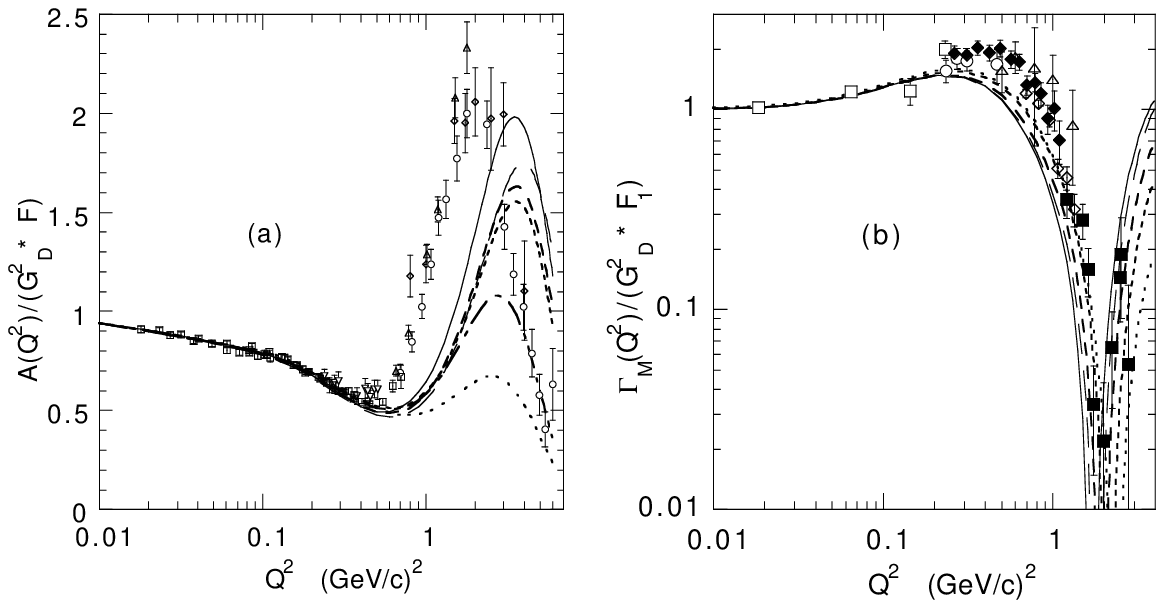,bbllx=8mm,bblly=212mm,bburx=0mm,bbury=282mm}
   
Figure 2. { (a) The ratio $A(Q^2)/(G_{D}^2 \cdot F)$ (with
$G_{D}=(1+Q^2/0.71)^{-2}$ and $F=(1+Q^2/0.1)^{-2.5}$) vs $Q^2$, obtained with
different $N-N$ interactions and the nucleon ff of Ref. \cite{Hoehler}. Solid
line: $RSC$ interaction; dashed line: $AV18$ interaction; dot-dashed line:
$Nijmegen1$ interaction; long-dashed line: $Nijmegen2$ interaction; short-dashed
line: $Nijmegen93$ interaction; dotted line: $CD-Bonn$ interaction. (b) As in
Fig. 1(a), but for the reduced form factor $\Gamma_M (Q^2) / ( G_D^2 \cdot F_1
)$  with $\Gamma_M (Q^2) = [G_M (Q^2) m_p / (\mu_d m_d)]^2$ and $F_1 = (1 +
Q^2/0.1)^{-3}$. The $Nijmegen1$ result is very similar to the $CD - Bonn$ one
and is not reported in this figure. (After Ref. \cite{LPS3}) }

\end{figure}

\begin{figure}[t]

\psfig{figure=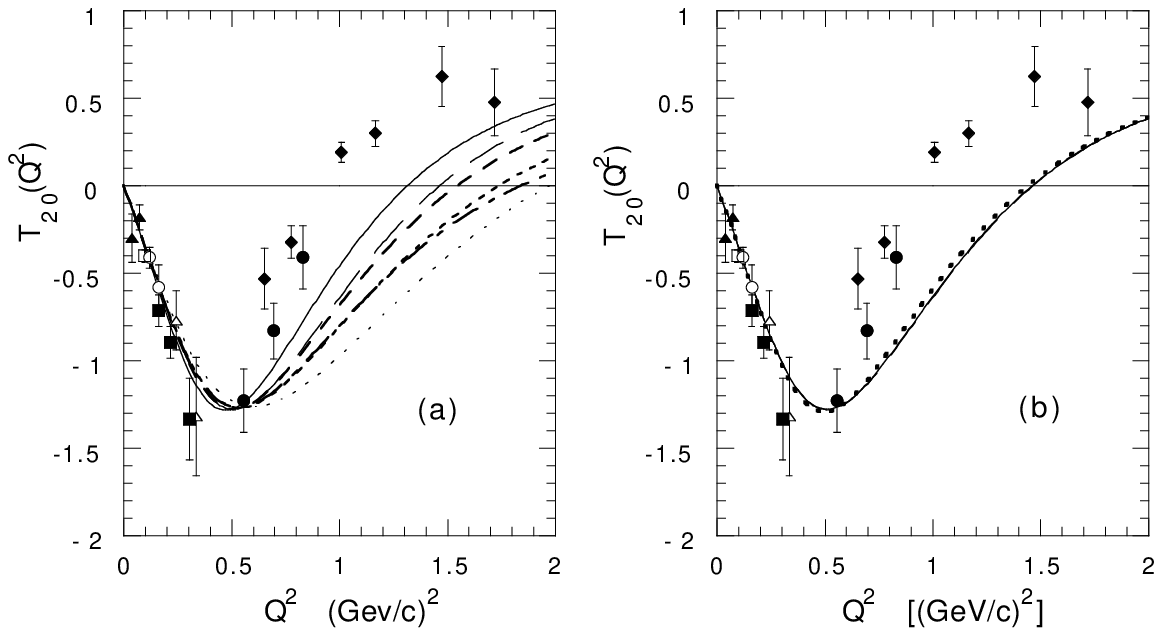,bbllx=11mm,bblly=208mm,bburx=0mm,bbury=282mm}

Figure 3. { (a) The tensor polarizations $T_{20}(Q^2)$ obtained with different
$N-N$ interactions and the nucleon ff of Ref. \cite{Hoehler}. The lines have 
the same meaning as in Fig. 2(a).
(b) The tensor polarization $T_{20}(Q^2)$ obtained with the nucleon ff models
of Ref. \cite{Mergell} (solid line), Ref. \cite{GK} (dotted line), and Ref.
 \cite{Hoehler} (dashed line).  (After Ref. \cite{LPS3}) }   

\end{figure}

In this section the deuteron electromagnetic form factors, obtained with our
current operator, are analyzed in detail by studying the dependence on different
realistic $N-N$ interactions and on different nucleon form factor models, but,
first of all, let us note the large differences between relativistic and non
relativistic calculations, even at low momentum transfer values ($Q^2 \simeq 0.2
- 0.3$ $(GeV/c)^2$) \cite{LPS3}. 

As shown in Figs. 2, 3 and 4, different interactions yield huge effects on
$A(Q^2)$ at $Q^2 \ge 1$ $(GeV/c)^2$, while for $B(Q^2)$ and the tensor
polarizations their effects are large already at $Q^2 \ge 0.5$ $(GeV/c)^2$ (see
also Ref. \cite{LPS3}). In Fig. 2 a suitable reduced plot as been used for 
$A(Q^2)$ and $B(Q^2)$ to allow a closer comparison between theoretical results
and experimental data. An interesting correlation between the S-state deuteron
momentum distribution and the deuteron form factors has been found. Indeed,
interactions which have essentially the same S-state momentum distribution,
$n_o(k)$, as the $AV18$ and $Reid93$ interactions, give the same results for
$A(Q^2)$, $B(Q^2)$ and the tensor polarizations (for this reason the curves
corresponding to $Reid93$ are not reported in the figures). Furthermore, as
shown in Fig. 5, the behaviour of $B(Q^2)$ and of the S-state deuteron momentum
distribution for corresponding interactions is similar. In order to
perform the comparison between the deuteron ff and S-state momentum
distribution on a quantitative basis, we report in Fig. 6 the position of the
minimum of $B(Q^2)$ and the position of the second zero of $T_{20}(Q^2)$
against the nonrelativistic S-state kinetic energy, $T_S$. A distinct linear
behaviour has been found: an higher value of $T_S$ moves the minimum of
$B(Q^2)$ and the second zero of $T_{20}(Q^2)$ to a lower momentum transfer
\cite{LPS3}$^-$\cite{LPS5}.

\begin{figure}[t]

\psfig{figure=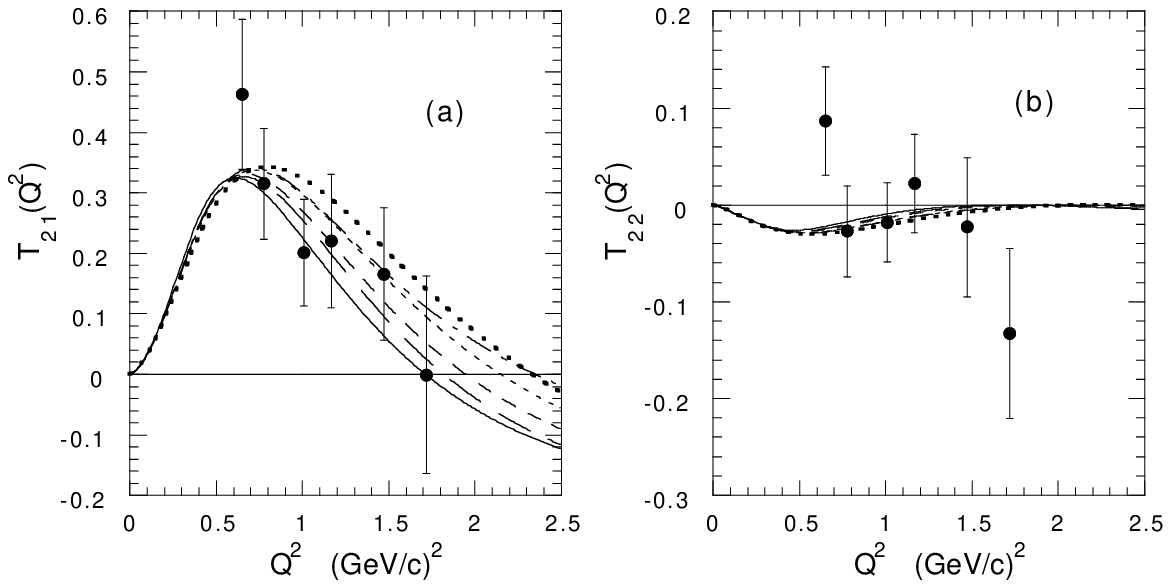,bbllx=10mm,bblly=210mm,bburx=0mm,bbury=282mm}

Figure 4. { The tensor polarizations $T_{21}(Q^2)$ (a) and $T_{22}(Q^2)$ (b),
obtained with different $N-N$ interactions and the nucleon ff of Ref.
\cite{Hoehler}. The lines have the same meaning as in Fig. 2(a). Experimental
data are from Ref. \cite{AbT}. (After Ref. \cite{Evora}) }   

\end{figure}

\begin{figure}[t]

\psfig{figure=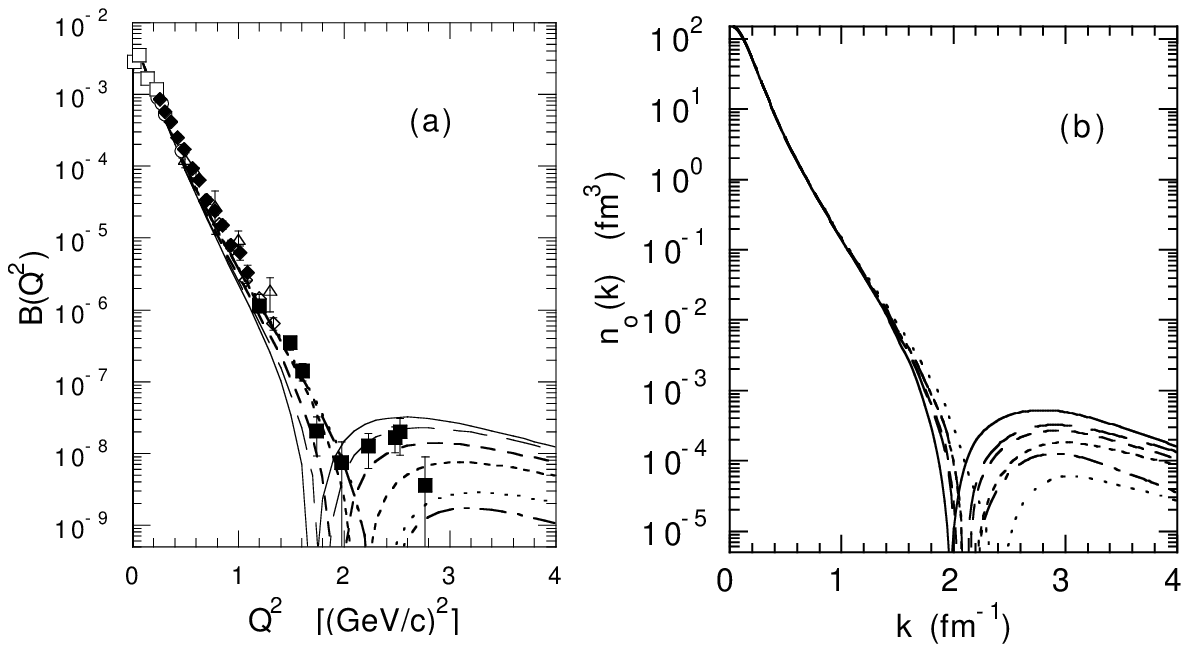,bbllx=10mm,bblly=204mm,bburx=0mm,bbury=280mm}

Figure 5. { (a) The deuteron ff $B(Q^2)$ obtained with different $N-N$
interactions and the nucleon ff of Ref. \cite{Hoehler}. The lines have the same
meaning as in Fig. 2(a). (After Ref. \cite{LPS3}) (b) The S-state deuteron
momentum distribution corresponding to different $N-N$ interactions. The lines
have the same meaning as in Fig. 2(a).}   

\end{figure}
Within an analysis which neglects two-body currents and isobar components, this
behaviour suggests the possibility of a description of the experimental data for
$T_{20}(Q^2)$ with interactions exhibiting high values of $T_S$ \cite{LPS3}. Let
us note that recent measurements of the $S-D$ mixing parameter, $\epsilon_1$
\cite{Ra}, point to a stronger tensor force than the one exhibited by the
interaction models we have analyzed. In turn, a stronger tensor force is
favoured by an high degree of locality, which yields significantly larger
kinetic energies and, in particular, larger values of $T_S$ \cite{Polls}.
 Then, one can argue that a $N-N$ interaction, able to reproduce the recent
measurements of $\epsilon_1$, could yield agreement between experimental and
theoretical values for $T_{20}(Q^2)$ and a minimum for $B(Q^2)$ slightly lower
than the value indicated by the available experimental data. Therefore, if the
new, more precise, measurement of $B(Q^2)$ which is planned at TJNAF will
show a lower value for the position of the minimum, both $T_{20}(Q^2)$ and
$B(Q^2)$ could be reproduced, without a relevant role of dynamical two-body
currents and isobar components.

\begin{figure}[t]

\psfig{figure=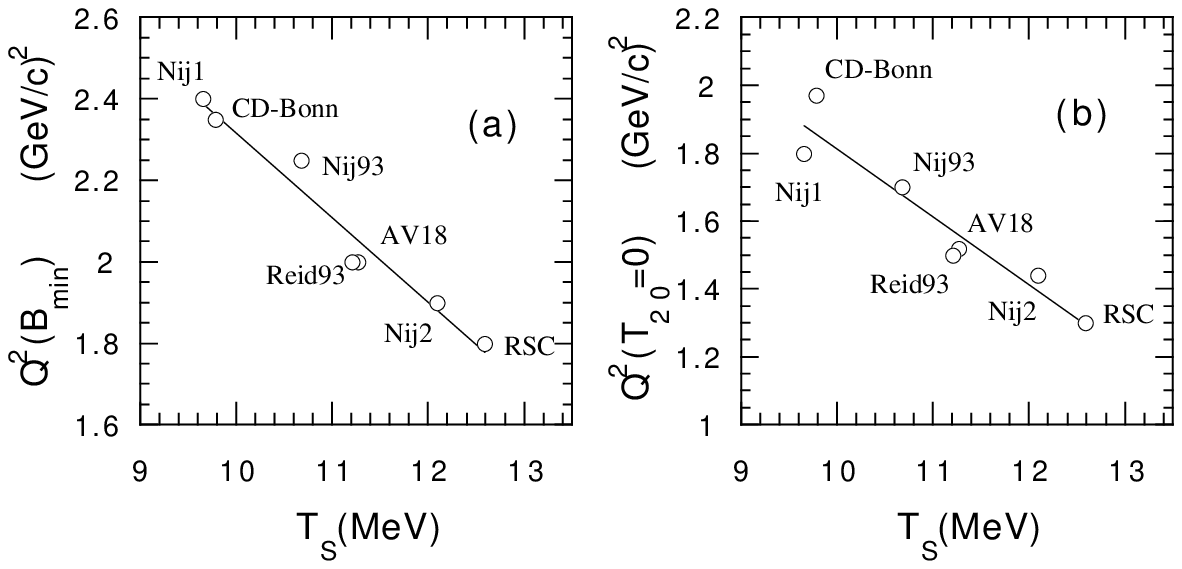,bbllx=8mm,bblly=209mm,bburx=0mm,bbury=282mm}

Figure 6. { (a) The position of the minimum of $B(Q^2)$, and (b) the position
of the second zero of $T_{20}(Q^2)$, corresponding to the nucleon ff of Ref.
\cite{GK}, vs the nonrelativistic S-state kinetic energy of the deuteron for
different realistic interactions. The experimental values of $Q^2(B_{min})$ and
$Q^2(T_{20} = 0)$ are $\simeq 1.8$ $(GeV/c)^2$ and $\simeq 1.0$ $(GeV/c)^2$,
respectively. (After Ref. \cite{LPS3}) }     

\end{figure} 

 Different nucleon ff models have very large effects on $A(Q^2)$, small effects
on $B(Q^2)$ and negligible effects on the tensor polarizations $T_{20}(Q^2)$,
$T_{21}(Q^2)$, $T_{22}(Q^2)$ (see Figs. 3(b), 7(a) and Ref. \cite{LPS3}). In
view of the large uncertainties in the experimental knowledge of the neutron
charge ff, we have tried to find a phenomenological description for the nucleon
charge ff, able to describe the deuteron elastic data for $A(Q^2)$ and, at the
same time, to reproduce the recent data for the ratio $G_E^p/G_M^p$
\cite{Perdri}. The proton and neutron magnetic ff have been fixed to the model
of Ref. \cite{GK} and the parametrization of the same model has been used for
the proton and neutron charge ff, but changing the values of the parameters. As
shown by the solid and long-dashed lines in Fig. 7(a), corresponding to the
$RSC$ and $Reid93$ interactions, respectively, an excellent agreement with the
experimental data for $A(Q^2)$ can be obtained, except for small differences in
the region around $Q^2=1$ $(GeV/c)^2$. The obtained $G^p_E(Q^2)$ goes exactly
through the data of Ref. \cite{Perdri}, while $G^n_E(Q^2)$ turns out to be
somewhat higher than the Gari-Kr\"{u}mpelmann model around $Q^2 = 0.8$
$(GeV/c)^2$ and smaller beyond. Similar results can also be obtained using other
realistic interactions. Changing the $N-N$ interaction a range of possible
values for $G^n_E(Q^2)$ could be identified. 

\begin{figure}[t]

\psfig{figure=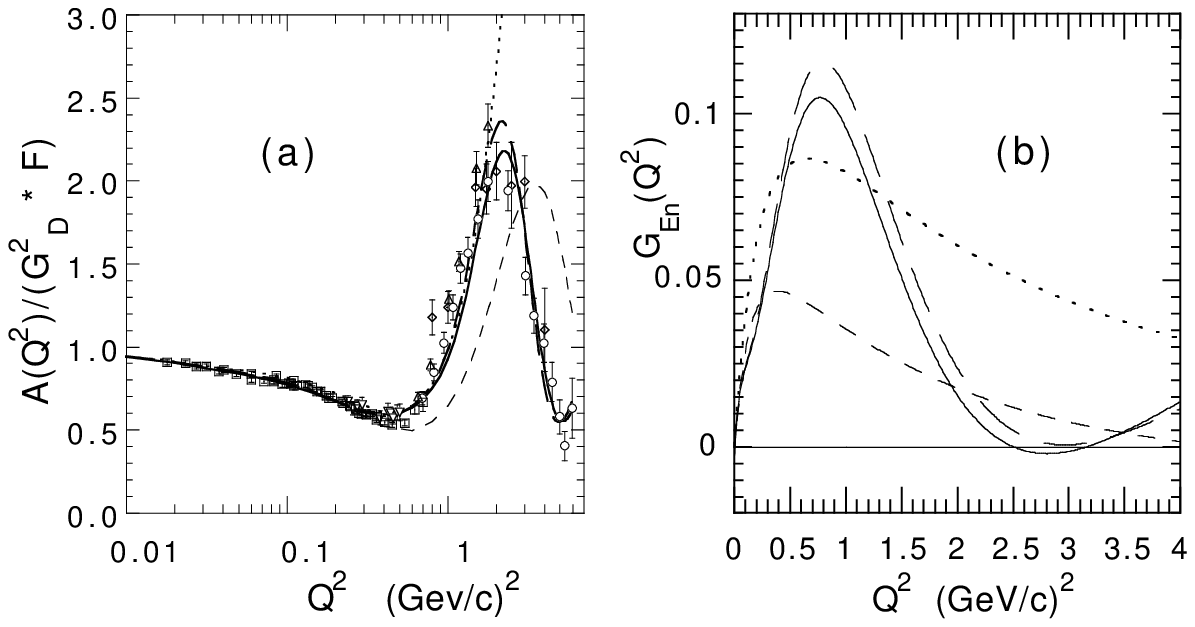,bbllx=8mm,bblly=200mm,bburx=0mm,bbury=280mm}

Figure 7. { (a) Solid and long-dashed lines: the ratio $A(Q^2)/(G_{D}^2 \cdot F)$
calculated with $RSC$ and $Reid93$ interactions, respectively, and fitted nucleon
charge ff (see text); dotted and short-dashed lines: the same as the long-dashed
line, but with the nucleon ff of Ref. \cite{GK} and Ref. \cite{Hoehler},
respectively. (After Ref. \cite{Taipei}) (b) The neutron charge form factor,
$G_E^n(Q^2)$. Solid and long-dashed lines are the fitted ff, corresponding to
the  $RSC$ and $Reid93$ interactions, respectively; dotted and short-dashed lines
are the neutron charge ff of Ref. \cite{GK} and Ref. \cite{Hoehler},
respectively.}

\end{figure}

However, these results on $G^n_E(Q^2)$ cannot be taken too seriously. Indeed,
the contributions of explicit two-body currents and of isobar components in the
deuteron wave function have to be thoroughly investigated, before drawing
definite conclusions.

\section{Deuteron Deep Inelastic Structure Function}

The Poincar\'e-covariant current operator can also be applied to the deep
inelastic electron scattering off nuclei \cite{LPS1,PSS1}.
In convolution models \cite{Coester} the deuteron DISF, in the Bjorken limit, is

\begin{eqnarray}
F_2^{D}(x) = 
\int\nolimits_{x}^{M_D/M} [ F_2^{p}(x/z) + F_2^{n}(x/z) ] f^{D}(z) dz  
\label{1}
\end{eqnarray}
where $x = Q^2/(2M \nu)$, while $M$ and $M_D$ are the nucleon and deuteron
masses, respectively. The distribution $f^D(z)$ describes the structure of the
deuteron system. In the usual approach \cite{CL,CSPS}, which makes use of the
impulse approximation within an instant-form framework, the distribution
$f^D(z)$ has the following expression 

\begin{figure}[t]

\psfig{figure=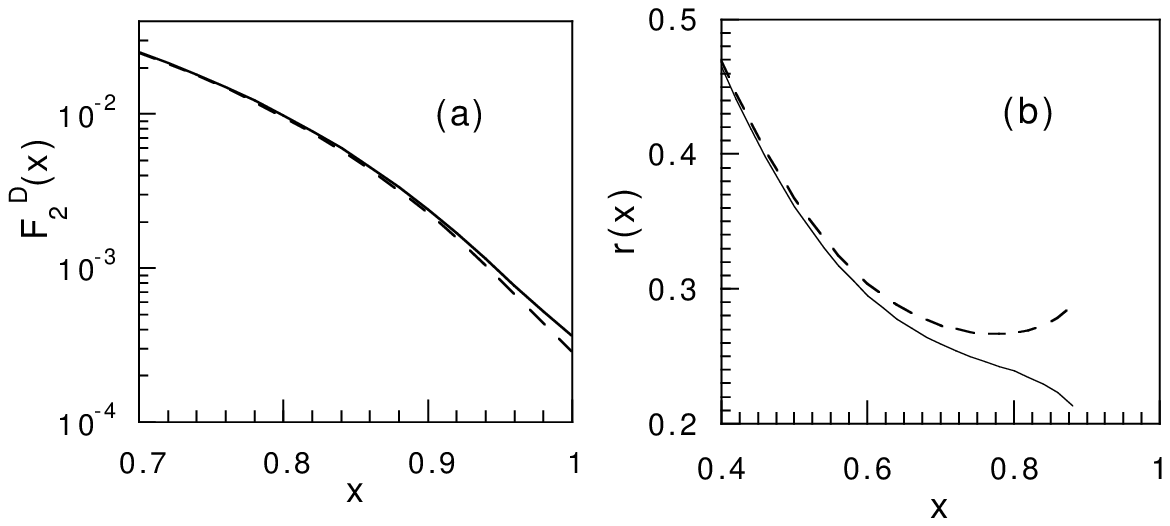,bbllx=8mm,bblly=223mm,bburx=0mm,bbury=282mm}

Figure 8. { (a) The deuteron DISF, $F_2^D(x)$, obtained from Eq. (\ref{1})
using the $AV18$ $N-N$ interaction and (b) the ratio $r(x)$ of neutron to proton
DISF, extracted by the recurrence relation (Eq. (\ref{4})) from simulated data
for the deuteron DISF (see text). The dashed and solid lines are obtained
using in the denominator of Eq. (\ref{4}) the expressions of Eqs. (\ref{2}) or
(\ref{3}) for the distribution $f^D(z)$, corresponding to the impulse
approximation within the instant-form approach or to our Poincar\'e-covariant
current operator, respectively. }     

\end{figure}
\begin{eqnarray} 
f^{D}_{IF}(z) = 
\int\nolimits d {\vec p} \; n^D( |{\vec p} |) \; \delta \left( z -
\frac {pq} {M\nu} \right) \; z \; C \;  
\label{2}
\end{eqnarray}
where $n^D( |{\vec p} |)$ is the nucleon momentum distribution in the deuteron,
$C$ a normalization factor, $q \equiv (\nu, {\vec q})$ the four-momentum
transfer, and $p$ the four-momentum of an off-mass shell nucleon, i.e., 
$p \equiv (p^0, {\vec p})$ with $p^0 = M_D - \sqrt{M^2 + |{\vec p}|^2}$.
Within the front-form Hamiltonian dynamics, using the Poincar\'e-covariant
current operator, the expression for the deuteron DISF is again given by Eq.
(\ref{1}), but the distribution $f^D(z)$ reads as follows \cite{PSS1} (see
also Ref. \cite{Coester}):  
\begin{eqnarray} 
f^{D}_{FF}(z) = 
\int\nolimits d {\vec k} \; n^D( |{\vec k} |) \; \delta \left( z -  {\xi}
\frac {M_D} {M} \right)   
\label{3}
\end{eqnarray}
where $ \xi = p^+/P^+ = 
[k_z + \sqrt{M^2 + |{\vec k} |^2} ]/ [2 \sqrt{M^2 + |{\vec k} |^2}] $. It has to
be stressed that, at variance with the instant-form dynamics, in the front-form
Hamiltonian dynamics the nucleons are always on their mass shell,
i.e. $p^2=M^2$. 
In Fig. 8(a) we report the deuteron DISF obtained with the nucleon momentum
distribution corresponding to the $AV18$ $N-N$ interaction, using the model of
Ref. \cite {Aubert} for the nucleon DISF. The difference between the results
obtained with the two different expressions for $f^{D}(z)$ is very small, except
at $ x \sim 1$.

In Ref. \cite{PSS} it has been proposed to extract the ratio
$r(x) = F_2^n(x)/F_2^p(x)$ from the experimental data for the deuteron DISF,
$F_2^{D exp}(x)$, by the following recurrence relation :
\begin{eqnarray}  
r^{(n+1)}(x) =
\frac{F_2^{D exp}(x) [1 + r^{(n)}(x) ] } 
{\int\nolimits_{x}^{M_D/M} [ 1 + r^{(n)}(x/z) ] F_2^{p}(x/z) f^{D}(z) dz } - 1 
\label{4} 
\end{eqnarray}
It was shown \cite{PSS} that a rapid convergence can be obtained, largely
independent on the assumed zero-order approximation, $r^{(0)}(x)$, up to $x
\leq 0.90$.

 In order to investigate the sensitivity of the extraction of $r(x)$ to
different expressions for $f^{D}(z)$ (cf. Eq. (\ref{2}) and (\ref{3})), we have
extracted $r(x)$ by simulating the data of the deuteron DISF, $F_2^{D exp}$,
through a theoretical evaluation. As a matter of fact, 
the deuteron DISF has been calculated by Eq. (\ref{1}) using for $f^{D}(z)$
the average of $f_{FF}^{D}(z)$ and $f_{IF}^{D}(z)$, corresponding to the $AV18$
$N-N$ interaction and the model of Ref. \cite {Aubert} for the nucleon DISF.
Then, Eq. (\ref{4}) has been applied, by inserting in the denominator
$f_{FF}^{D}(z)$ or $f_{IF}^{D}(z)$, alternatively, always corresponding to 
the $AV18$ interaction. As shown in Fig. 8(b), the differences in the extracted
functions $r(x)$ become quite sizeable at $x \ge 0.7$. Moreover we have shown
that the differences in the extracted values of $r(x)$ are of the same order if
$f_{FF}^{D}(z)$ or $f_{IF}^{D}(z)$ are used to generate the simulated data,
instead of their average value.  Therefore a Poincar\'e-covariant treatment of
relativity appears to be important in the extraction of the neutron deep
inelastic structure function at high $x$. An analysis of the experimental data
for $F_2^{D}$ will be performed in Ref. \cite{PSS1}.

\section{Conclusions}
Deuteron elastic form factors and deep inelastic structure functions have
been studied within the front-form Hamiltonian dynamics, using a
Poincar\'e-covariant current operator for bound systems of interacting
particles, which satisfies parity and time-reversal covariance, as well as
Hermiticity and current conservation. The current is built up from the free one
in the Breit reference frame where three-momentum transfer of the interacting
system is directed along the spin quantization axis, $z$. This current allows
one to obtain elastic and transition form factors, as well as deep inelastic
structure functions, with no ambiguity. 

The Poincar\'e-covariant relativistic approach has been shown to have a big
impact with respect to nonrelativistic calculations, both in the elastic and
deep inelastic case and even at low momentum transfer values.

As far as the elastic ff are concerned, the usual disagreement between
theoretical and experimental values for $\mu _d$ and $Q_d$ is largely removed
and therefore contributions beyond impulse approximation are expected to play a
minor role at $Q^2 \rightarrow 0$.
Large effects have been found from different nucleon form factor models and
different realistic $N-N$ interactions. It is especially noteworthy an
interesting correlation between the deuteron S-state kinetic energy, $T_S$, and
the positions of the minimum of $B(Q^2)$ and of the second zero of
$T_{20}(Q^2)$. 

We stress that, when comparing the theoretical results of different approaches
with the experimental data for the deuteron ff, all the observables should be 
considered at the same time, before drawing definite conclusions on the quality
of the various approaches. Furthermore, the possible different $N-N$
interactions and nucleon ff models should be carefully investigated.

Our next step will be to study within our approach the role of contributions
from explicit two-body current and from isobar configuration in the deuteron
wave function. 

As far as the deep inelastic structure function is concerned, the relevance of
a Poincar\'e-covariant relativistic approach has been clearly shown in the
extraction of the neutron DISF from the deuteron DISF. Even if the effects in
the calculation of $F_2^D(x)$ are small, the consequences of
Poincar\'e-covariance in the extraction of the neutron DISF become
important at high $x$, especially in view of the experiments planned at TJNAF
for measuring $F_2^n(x)$ up to $x = 0.83$ \cite{MP}.


\section*{Acknowledgments}
The authors are highly indebted to S. Scopetta for his kind help in obtaining
the deuteron deep inelastic structure function and in the extraction of the
neutron deep inelastic structure function.

\end{document}